\documentclass[aip,jcp,reprint,amsmath,amssymb,floatfix]{revtex4-2}
\usepackage{graphicx,booktabs,bm,xcolor}
\usepackage[colorlinks=true,allcolors=blue]{hyperref}
\graphicspath{{figures/}}
\newcommand{\dd}{\mathrm{d}}
\begin{document}
\title{Molecular Photoionization Spectra from Signed Dyson Orbitals and Anharmonic Nuclear Motion}
\author{Vincenzo Barone}
\email{vincebarone52@gmail.com}
\affiliation{INSTM, National Interuniversity Consortium for Materials Science and Technology, via G. Giusti 9, 50121 Firenze, Italy}
\date{\today}
\begin{abstract}
Molecular photoelectron bands couple changes in nuclear structure to the energy-dependent scattering of the departing electron. Band intensities require both nuclear transition moments and electronic continuum amplitudes. We present SPARKS (Scattering and Photoionization: Amplitudes, Response Kernels and Spectra), a molecular photoionization approach that connects signed Dyson orbitals to complex molecular dipole amplitudes and harmonic or anharmonic nuclear transition moments. Compact B-spline and finite-element discrete-variable representations support spectral and sparse outgoing-boundary solvers. Independent molecular checks reproduce a reference H$_2^+$ spectrum and CO continuum amplitudes, while analytic hydrogen photoionization tests establish the continuum normalization. For H$_2$O/H$_2$O$^+$, electronic derivatives activate a Franck--Condon-dark fundamental and enter the same nuclear operator used in harmonic, perturbative and internal-coordinate variational calculations. Furan extends the nuclear treatment to all 21 vibrational coordinates in a selected configuration-interaction window, including low-lying out-of-plane states. Four epichlorohydrin ionization channels show how orbital mixing reshapes cross sections and angular distributions at similar Dyson norms. Energy conservation is imposed separately for every vibronic transition at fixed photon energy. The resulting spectra retain vibrational assignments and electronic interference, connecting weak-band activation, anharmonic spacings and photoelectron angular distributions within one physical description.
\end{abstract}
\maketitle
\raggedbottom
\section{Introduction}
Photoelectron spectroscopy identifies ionization channels, measures molecular structural changes and follows the electronic character of reacting or excited molecules.\cite{Reid2003,Stolow2004} Vibrational progressions encode the displacement and reshaping of the ionic potential surface, while their intensities also depend on the outgoing electron. As photon energy changes, molecular scattering modifies both the magnitude and phase of the electronic transition amplitude. A weak vibrational band can consequently probe the coordinate dependence of the continuum operator as strongly as the nuclear potential.

This observation motivates an explicit connection between electronic-continuum calculations and vibrationally resolved spectra. The Dyson orbital collects the overlap between the neutral many-electron wavefunction and a specified ionic state. Its squared norm measures the weight of that removal channel, whereas its spatial structure and its contraction with the continuum determine the dipole amplitude.\cite{Oana2007,Ortiz2020} Plane- and Coulomb-wave treatments provide useful analytic descriptions of the departing electron, with their relative accuracy depending on the residual ionic field.\cite{Oana2009,Gozem2015} Established calculations combine correlated Dyson orbitals with B-spline density-functional continuum states.\cite{Tenorio2022,Tiresia2024} Multielectron R-matrix methods provide a complementary route to correlated scattering and photoionization,\cite{Tennyson2010,UKRmol2020} alongside variational scattering formulations based on the Schwinger principle.\cite{Lucchese1980} Vibrationally resolved photoionization studies have connected departures from the Condon approximation to molecular scattering, interference and diffraction.\cite{Plesiat2012} Engin and co-workers combined three-dimensional anharmonic water wavefunctions with continuum dipoles evaluated over the nuclear geometry space.\cite{Engin2019} Their work establishes full-dimensional water photoionization as a physical reference for the present hierarchy. Recent TIRESIA calculations on epichlorohydrin show how correlated orbital mixing changes partial cross sections and angular anisotropies.\cite{Schio2026} Vibrationally resolved water calculations with a correlated multichannel continuum also reveal mode-dependent attosecond delays.\cite{Pranjal2025} These developments motivate a consistent connection between the full electronic transition amplitude and assigned nuclear states. The choice of electronic model and the numerical representation of the continuum remain separate decisions.

The nuclear problem has an analogous hierarchy. Harmonic surfaces permit efficient time-independent (TI) transition assignments and time-dependent (TD) profiles, including the mixing of normal coordinates between electronic states.\cite{Doktorov1977,Vibronic2026} Perturbative corrections can describe moderately anharmonic levels, and variational nuclear wavefunctions provide direct control where mode coupling or weak transitions require greater accuracy.\cite{Christiansen2007,Bowman2008} Vibrational self-consistent-field treatments and coupled product-state expansions provide established routes to these anharmonic states.\cite{Carter1997} A useful interface must preserve complex electronic amplitudes, their coordinate derivatives and a consistent nuclear measure across this hierarchy.

Time-dependent nuclear motion offers a complementary description through spectral correlation functions. Multiconfigurational wavepacket propagation treats coupled quantum dynamics,\cite{Meyer1990} while Gaussian wavepacket approaches provide semiclassical approximations.\cite{Heller1975} Extensions beyond the Condon approximation and to finite temperature include mode coupling and coordinate-dependent optical amplitudes.\cite{Patoz2018,Begusic2020} Connecting such nuclear descriptions to molecular ionization requires the complex, electron-energy-dependent continuum operator.

The spectroscopic questions addressed here are therefore threefold: which weak bands arise from the electronic transition operator, how anharmonic motion changes their positions and intensities, and how the departing electron redistributes intensity as its energy changes. We introduce SPARKS (Scattering and Photoionization: Amplitudes, Response Kernels and Spectra) to answer these questions with a common complex transition operator. Its principal contributions are phase-consistent derivatives of correlated Dyson--continuum amplitudes, a shared signed-moment interface across harmonic, perturbative and variational nuclear descriptions, and energy conservation imposed separately for each vibrational transition. These features connect state assignments and angular distributions to the same electronic operator. Independent continuum representations distinguish numerical convergence from the choice of molecular potential.

Water provides a direct test of continuum-induced dark-band activation and anharmonic ionic spacings against experiment. Furan extends the treatment to coupled motion in all 21 vibrational coordinates and to weak out-of-plane bands in a specified low-energy window. Epichlorohydrin tests the complementary electronic question: how changes in removal-orbital shape redistribute strength among four ionic channels even when their Dyson norms are similar. The methodology first connects electronic scattering to nuclear transition moments; the results then follow the progression from independent continuum tests to assigned molecular spectra. Numerical settings, convergence sequences and reproduction details are collected in the Supporting Information (SI).

\begin{figure}[t]
\includegraphics[width=\linewidth]{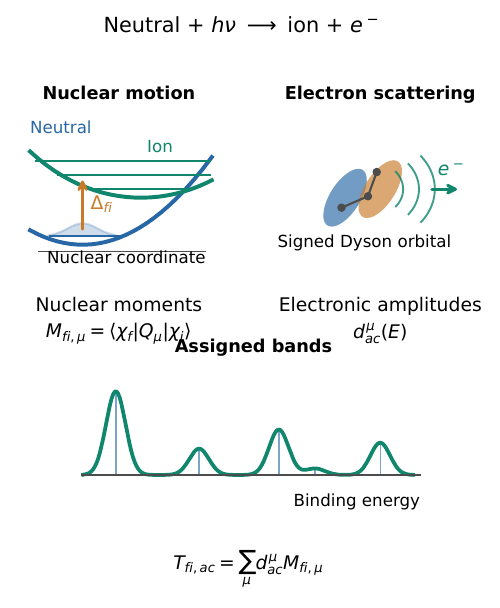}
\caption{Schematic connection between nuclear transition moments, electronic continuum amplitudes and assigned photoelectron spectra. Curves are illustrative; $Q_0=1$ denotes the constant operator.}
\label{fig:physical-scheme}
\end{figure}

Figure~\ref{fig:physical-scheme} summarizes the physical connection between nuclear reorganization, molecular electron scattering and the observed bands.

\section{Continuum and nuclear methodology}
\subsection{Dyson orbitals and electronic inputs}
For a neutral state $\Psi_N$ and ionic state $\Psi_{N-1}^{I}$, the spin-resolved Dyson orbital is
\begin{equation}
 \phi_D^I(x)=\sqrt{N}\int\dd X\,\Psi_{N-1}^{I*}(X)\Psi_N(x,X).
\end{equation}
Here $X=(x_2,\ldots,x_N)$ includes the remaining spin coordinates. The dipole amplitudes at nuclear coordinates $Q$ are
\begin{equation}
 d_{a c}^{I}(E;Q)=\langle\psi_{E c}^{I(-)}(Q)|r_a|\phi_D^I(Q)\rangle,
 \label{eq:dipole}
\end{equation}
where $a$ is Cartesian and $c$ labels the asymptotic electron channel. The continuum is normalized to $\delta(E-E')$. The signed Dyson coefficients, including their norm, enter Eq.~\eqref{eq:dipole} directly, so each channel weight is included once.

A common electronic-data interface reads \href{https://gaussian.com/}{Gaussian} formatted results, preserving atomic-orbital (AO) conventions, density matrices and molecular frames. Complete signed electron-propagator vectors are paired with the orbital data from the same calculation. Their orbital ordering and phase are preserved. Rigid AO rotations and cross-geometry overlaps follow the same Dyson state across displaced geometries, with the original electronic density and charge retained.

\subsection{Finite-space outgoing solution}
The central radial expansion uses compact B-splines multiplied by real spherical harmonics, following established continuum B-spline representations.\cite{Bachau2001} Atomic-centred spline blocks enrich the region near individual nuclei. Real local channel potentials generate a Bloch Hamiltonian $H$, overlap $S$, boundary columns $B$ and dipole source columns $D$. With $HU=SU\epsilon$, the spectral R matrix is\cite{Tennyson2010}
\begin{equation}
 R(E)=\tfrac12 B^{T}U(\epsilon-E)^{-1}U^{T}B.
\end{equation}
All eigenpairs are retained. For outgoing waves $h=G+iF$ and logarithmic derivative $q=h'h^{-1}$, the boundary system is
\begin{equation}
 [1-Rq]u=R(F'-qF).
 \label{eq:boundary}
\end{equation}
At an inner-region eigenvalue, a direct outgoing solve in the modal space evaluates the finite physical solution without shifting the energy.

The alternative sparse solver evaluates the corresponding full-space equation,
\begin{equation}
 [H-ES-\tfrac12 BqB^T]C=\tfrac12 B(F'-qF),
 \label{eq:sparse}
\end{equation}
and contracts $D^TC$. Exact radial support zeros are retained through assembly. The spectral solver amortizes its preparation across an energy scan, while the sparse solver avoids complete diagonalization and releases each factorization after use. Both implement the same boundary and source conventions.

Atomic and central angular integration are refined independently of the retained angular channels. The enriched overlap is required to be positive before constructing the electronic field and is checked again in the complete continuum assembly. The SI gives a local Schur-complement check that resolves the atomic contribution without a complete overlap diagonalization.

A finite-element form of the discrete-variable representation (DVR), denoted FE-DVR, provides a second radial discretization.\cite{Light1985,Rescigno2000} Normalized Gauss--Lobatto cardinal functions share bridge functions at element boundaries. The origin degree of freedom is removed, and the outer endpoint remains open. Lebedev angular quadrature is available alongside the product angular rule.\cite{Lebedev1999} These choices change the numerical representation without changing the declared molecular potential.

\subsection{Local field and molecular asymptote}
The local molecular model contains the nuclear potential, the exact AO Hartree field and the spin-unpolarized van Leeuwen--Baerends (LB94) exchange--correlation potential.\cite{LB1994} The density and analytic spatial gradient are evaluated from the imported AO state. The local-density part uses the LibXC implementation.\cite{LibXC2018} Gaussian AO values and molecular integrals are evaluated with established electronic-structure libraries.\cite{PySCF2020}

A finite Gaussian expansion has a Gaussian density tail. Inserting that tail into LB94 gives a different ultimate asymptote from inserting an exponential molecular density. The adopted molecular model therefore joins the exchange--correlation contribution to $-1/r$ with an explicit smooth switch between 10 and 15 bohr. Nuclear and Hartree multipoles are preserved. The switch defines the electronic model; its sensitivity is examined separately from discretization convergence. The static model uses local exchange and treats each ionic electronic channel separately. Angular partial-wave coupling is retained within that channel. Nonlocal exchange and coupling between distinct ionic electronic states require an extended electronic model, as illustrated by multichannel water calculations.\cite{WaterXchem2023}

The exterior field is propagated through complete channel matrices. A dipole--Coulomb reference diagonalizes
\begin{equation}
 A_{cc'}=l_c(l_c+1)\delta_{cc'}+2D_{cc'},
\end{equation}
where $D$ is the molecular dipole coupling. The resulting radial equations retain negative centrifugal eigenvalues. This treatment removes the artificial finite-radius truncation of the permanent dipole while restoring the original channel convention after propagation.

\subsection{Electronic response and the spectral model}
The retarded electronic response includes the induced Hartree and exchange--correlation field through continuum time-dependent density-functional theory (TDDFT).\cite{Stener2005,Decleva2022} Its formulation and atomic and molecular verification are collected in the SI. All vibrational spectra in this article use the static local field specified above; electronic screening is tested as a separate extension.

\subsection{Nuclear transition operator}
A linear expansion in declared dimensionless nuclear coordinates gives
\begin{equation}
 d_{ac}(E;Q)=d^0_{ac}(E)+\sum_j d^j_{ac}(E)Q_j.
 \label{eq:ht}
\end{equation}
For nuclear states $i,f$, define $M_{fi,0}=\langle\chi_f|\chi_i\rangle$ and $M_{fi,j}=\langle\chi_f|Q_j|\chi_i\rangle$. Their contraction yields
\begin{equation}
 T_{fi,ac}(E)=\sum_\mu d^\mu_{ac}(E)M_{fi,\mu}.
 \label{eq:nuclear}
\end{equation}
Summing the electronic channels first forms the small Hermitian matrix
$W_{\mu\nu}(E)=\sum_{ac}d^\mu_{ac}(E)d^{\nu*}_{ac}(E)$.
Its constant, derivative and mixed blocks give Franck--Condon (FC), Herzberg--Teller (HT) and interference contributions. The contraction retains every supplied electronic channel.

Electron-propagator calculations supply both removal energies and correlated Dyson vectors.\cite{Ortiz2013,Ortiz2020} In an orthonormal orbital basis, the removal part of the retarded Green function has the pole expansion
\begin{equation}
 G^-_{pq}(z)=\sum_I\frac{X_p^I X_q^{I*}}{z+\mathrm{IP}_I},
 \qquad P_I=\sum_p|X_p^I|^2,
\end{equation}
where $\mathrm{IP}_I$ is an ionization energy and $X^I$ is the signed Dyson vector. Gaussian renormalized second-order electron-propagator (NR2) calculations provide these quantities.\cite{NR2} Nuclear derivatives follow the same removal channel by cross-geometry orbital overlap. The continuum derivative includes changes in the complete Dyson vector and scattering field; HT intensity depends on the complete amplitude derivative, including its phase.

Spatial symmetry reduces the displaced electronic set. For a reflection $R$ that maps $+Q_j$ onto $-Q_j$, the dipole amplitudes satisfy
\begin{equation}
 d(-Q_j)=\eta_I R\,d(+Q_j)U(R)^{\mathsf T},
 \label{eq:reflection}
\end{equation}
where $U(R)$ acts on the real spherical channels and $\eta_I=\pm1$ is the reference Dyson character. Transforming amplitudes preserves the odd HT response of an out-of-plane mode even when its two scalar pole strengths are equal.

The same $M$ interface accepts harmonic overlaps, second-order vibrational perturbation theory (VPT2), or variational overlaps. Harmonic TI and TD share an analytic sum rule. The variational water calculation instead contracts quadrature wavefunctions from a full curvilinear $J=0$ Hamiltonian in the two O--H distances and H--O--H angle. The physical volume connection and all mixed kinetic terms are included, consistently with the coordinate-dependent quantum kinetic operator.\cite{Podolsky1928} Analytic Zoomable Asymptotically-Correct Force Field (ZAFF) potential surfaces supply the dense quadrature values. Their fit coefficients and electronic acquisition set are frozen before nuclear basis convergence.

Table~\ref{tab:model-guide} connects the required physical inputs to the observables and nuclear approximations. Each nuclear treatment supplies the signed moments in Eq.~\eqref{eq:nuclear}; the electronic amplitude and its coordinate derivatives retain the common frame and phase convention.
\begin{table*}[t]
\caption{Physical inputs and outputs of the electronic and nuclear descriptions. The nuclear approximation is chosen according to mode coupling and the spectral window of interest.}
\label{tab:model-guide}
\begin{ruledtabular}
\small
\setlength{\tabcolsep}{3pt}
\renewcommand{\arraystretch}{1.15}
\begin{tabular}{p{.17\textwidth}p{.43\textwidth}p{.32\textwidth}}
Description & Required inputs & Obtained quantities\\\hline
Electronic continuum & Geometry, AO density and basis, signed Dyson orbital, removal energy; phase-followed displaced data for HT & Complex continuum dipoles, their coordinate derivatives and angular channels\\
Harmonic nuclear & Two equilibrium structures and Hessians; coordinate transformation and displacement & Signed harmonic moments, TI assignments and TD profiles\\
Anharmonic perturbative & Quadratic, cubic and quartic fields in a consistent coordinate representation & Perturbative energies and signed nuclear moments\\
Anharmonic variational & Analytic potentials or a specified local force field, kinetic operator and nuclear basis & Coupled nuclear states and signed transition moments\\
Fixed-photon spectrum & Electronic amplitudes, nuclear moments, photon energy and channel multiplicity & Line-specific cross sections, FC/HT interference and angular anisotropies\\
\end{tabular}
\end{ruledtabular}
\end{table*}

\subsection{Fixed-photon spectra and normalization}
For a nuclear transition with binding energy $\Delta_{fi}$, energy conservation requires
\begin{equation}
 E_{fi}=\omega-\Delta_{fi}>0.
\end{equation}
The isotropic one-photon partial cross section is
\begin{equation}
 \sigma_{fi}(\omega)=\frac{4\pi^2\alpha\omega}{3}
 p_i g_I\sum_{ac}|T_{fi,ac}(E_{fi})|^2,
 \label{eq:crosssection}
\end{equation}
in Bohr areas. Here $p_i$ is the initial population and $g_I$ is an explicitly supplied channel multiplicity. Energy normalization already includes the continuum density of states; no additional electron momentum factor occurs. For the closed-shell water and furan spatial removal poles, two orthogonal ionic spin components give $g_I=2$. A megabarn (Mb) equals $10^{-22}$~m$^2$.

Continuum amplitudes are evaluated for open transitions and interpolated within the computed electron-energy interval, using a common channel and phase convention before contraction and squaring. Direct intermediate-energy evaluations test this interpolation independently. Instrumental Gaussian broadening is applied after transition-specific cross sections have been obtained. A profile $S(x)=\sum_f s_f g_\sigma(x-x_f)$ uses a unit-area Gaussian $g_\sigma$; its ordinate is integrated line strength per unit energy or wavenumber. Normalized profiles are identified separately from absolute cross-section densities.

\section{Computational protocol}
The continuum checks use the H$_2^+$ and CO examples distributed with the established TIRESIA program, matching occupations, Coulomb phases, fitting radii and symmetrization conventions.\cite{Tiresia2024} Water uses the revised dispersion-corrected, spin-component-scaled double hybrid revDSD-PBEP86-D3 with the jun-cc-pVTZ basis for its neutral and cationic nuclear fields,\cite{revDSD2019} and NR2/jun-cc-pVTZ for the correlated Dyson orbital.\cite{NR2} Coordinate frames, state identities and signed Dyson coefficients are preserved throughout.

For water, a 20-bohr inner sphere contains degree-seven splines, 80 nominal radial intervals and nuclear-radius integration cuts. The production angular space has $l\leq7$, or 64 real spherical channels. Exterior propagation extends to 120 bohr. The linear HT stencil uses existing displacements of $\pm0.02$ in each neutral normal coordinate. A five-point stencil at $l\leq7$ checks its step dependence on the production electron-energy grid; Fig.~\ref{fig:conditional} retains the separate five-point $l\leq5$ fixed-electron-energy example. Every geometry shares the spline trial space, with moving nuclear radii used only as quadrature cuts.

The internal-coordinate water surfaces use inverse-distance and angular analytic functions and previously acquired energies, gradients and Hessians. The final nuclear domain is $R_1,R_2\in[0.60,1.60]$~\AA\ and $\theta\in[40^\circ,178^\circ]$. Each coordinate has 28 sine functions and 60 Gaussian quadrature nodes. Twelve ionic levels and the neutral ground state supply Eq.~\eqref{eq:nuclear}. Basis, quadrature, domain and continuum checks are reported separately in the SI.

The molecular spectra start from the neutral vibrational ground state at zero temperature. The SI specifies the computational resources and numerical settings.

\section{Results and discussion}
\subsection{Independent normalization and molecular reference tests}
Table~\ref{tab:reference} summarizes independent tests of the continuum calculation. Hydrogen provides an analytic cross-section reference, including the source and continuum normalization. H$_2^+$ then tests independent multicentre assembly and dipoles over a complete energy scan. CO checks all three continuum symmetries with common operators; separate CO tests examine construction of the local electronic field.
\begin{table}[tbp]
\caption{Continuum checks. Errors are maximum observed relative differences unless an energy unit is shown.}
\label{tab:reference}
\begin{ruledtabular}
\begin{tabular}{lr}
Test & Difference\\\hline
Hydrogen analytic cross section & $2.4\times10^{-10}$\\
H$_2^+$ bound energy / hartree & $1.2\times10^{-11}$\\
H$_2^+$ independent spectrum, 160 energies & $2.80\times10^{-8}$\\
CO complex amplitudes, 150 energies & $5.95\times10^{-8}$\\
CO summed strength, common operators & $9.53\times10^{-9}$\\
CO matched-field central $H$ / hartree & $1.37\times10^{-12}$\\
\end{tabular}
\end{ruledtabular}
\end{table}
Independent angular integration reproduces the reference anisotropy parameter $\beta$ within $4.5\times10^{-8}$ for H$_2^+$ and CO. This checks Coulomb phases and angular contraction separately from the outgoing solver. Matched-potential and independent Coulomb-integral checks also establish the local-field construction; their detailed residuals are given in the SI.

The retarded response supplies a second normalization check. For atomic hydrogen, the absorptive susceptibility $-\operatorname{Im}\chi_{zz}/\pi$ agrees with the outgoing dipole strength within $2.67\times10^{-11}$ relative over three electron energies. This comparison retains both Sternheimer frequency terms and an explicit Coulomb boundary self-energy. The SI gives independent virtual-state and Whittaker-function controls.

\subsection{Continuum derivatives and a dark water fundamental}
A weak photoelectron band can originate in a changing electronic amplitude even when the removal-channel weight is stationary. The water calculation follows the complete signed Dyson orbital across nuclear displacements, retaining both this change and the response of the scattering field. The SI reports orbital-overlap and finite-difference checks.

\begin{figure}[tbp]
\includegraphics[width=\linewidth]{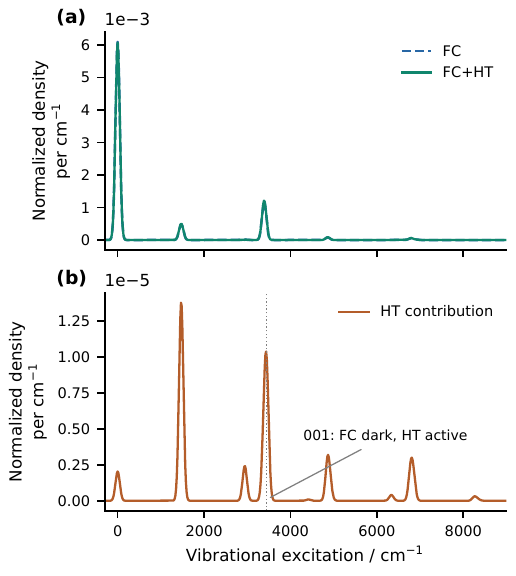}
\caption{Water at electron energy 0.20 hartree: total FC and FC+HT profiles (top) and HT contribution (bottom), using the five-point $l\leq5$ stencil.}
\label{fig:conditional}
\end{figure}

Figure~\ref{fig:conditional} shows how the continuum operator modifies the nuclear progression. The antisymmetric fundamental is dark in the FC sector and gains HT intensity, despite a vanishing first derivative of the pole strength. This distinction illustrates why a scalar quasiparticle weight cannot replace the signed continuum dipole and its derivatives. At electron energy 0.20 hartree the integrated HT/FC ratio is 0.00488 for the five-point $l\leq5$ operator shown in the figure. The separate linear $\pm0.02$, $l\leq7$ calculation gives 0.00506. These values identify the angular space and stencil together.
The harmonic TI sum over 680 transitions exhausts the FC/HT sum rule, and its TD correlation agrees to numerical precision; quantitative residuals are supplied in the SI. These checks concern a fixed-electron-energy operator; Eq.~\eqref{eq:crosssection} subsequently evaluates a different electron energy for each nuclear line.

\subsection{Anharmonic nuclear structure and experimental spacings}
\begin{figure}[tbp]
\includegraphics[width=\linewidth]{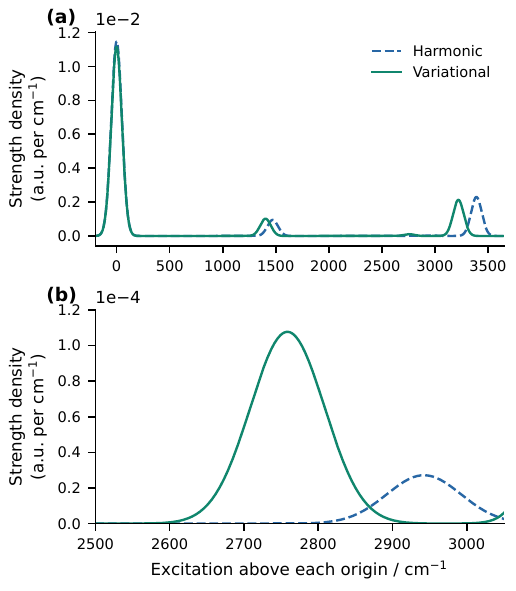}
\caption{Water harmonic and variational profiles at electron energy 0.20 hartree: five assigned bands (top) and bending overtone (bottom). Strength densities are in $(a_0^2/E_h)$ per cm$^{-1}$, relative to each origin.}
\label{fig:internal}
\end{figure}

\begin{figure}[tbp]
\includegraphics[width=\linewidth]{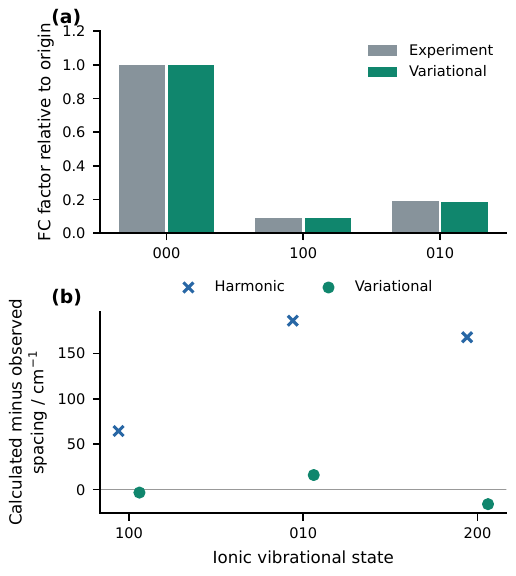}
\caption{Water comparison with experiment: (a) nuclear FC factors relative to the origin; (b) deviations from measured ionic spacings.}
\label{fig:experiment}
\end{figure}

Figure~\ref{fig:internal} and Table~\ref{tab:water} compare the harmonic and full internal-coordinate variational water calculations. Bending and stretching excitations shift substantially, while the relative effect on intensities depends on band character. The overtone is particularly sensitive to nuclear coupling and the chosen potential representation.
\begin{table}[tbp]
\caption{Water ionic spacings (cm$^{-1}$). Occupations are ordered bend, symmetric stretch, antisymmetric stretch. Independent three-dimensional (3D) theory is from Engin \emph{et al.}\cite{Engin2019}; experiment is from Reutt \emph{et al.}\cite{Reutt1986}}
\label{tab:water}
\begin{ruledtabular}
\setlength{\tabcolsep}{3pt}
\begin{tabular}{lrrrr}
Band & Harmonic & ZAFF/DVR & 3D theory & Experiment\\\hline
100 &1471.20&1403.42&1445.3&1406.63\\
010 &3391.25&3221.34&3173.8&3205.25\\
001 &3450.22&3268.89&3214.1&---\\
200 &2942.39&2758.59&2849.6&2774.55\\
\end{tabular}
\end{ruledtabular}
\end{table}
The independent three-dimensional calculation uses complete-active-space potential surfaces and includes coupling among all three vibrations. Its spacings provide a comparison of nuclear potential models; the SI gives the source quantum-number convention and unit conversion. The variational spacing errors for the measured bending fundamental, symmetric stretch and bending overtone are approximately $-3.2$, $+16.1$ and $-16.0$~cm$^{-1}$. Figure~\ref{fig:experiment} also compares nuclear FC factors with the photoelectron values tabulated by Botter and Rosenstock.\cite{Botter1969} The origin, bending and symmetric-stretch factors are 0.7499, 0.0668 and 0.1389, against experimental estimates 0.757, 0.069 and 0.143. These measurements test the Condon nuclear sector separately from the continuum contribution.
Independent basis, quadrature and domain refinements stabilize these five assigned bands; the largest spacing change on extending the physical domain is 0.0151~cm$^{-1}$. The twelve retained ionic states carry about 99.1\% of the fixed-energy linear-operator sum rule.

A matched-Hamiltonian test isolates the perturbative approximation by using the same rectilinear quartic potential in strict wavefunction VPT2 and a local variational calculation. The bending-overtone strengths are 0.003258 and 0.000634~$a_0^2/E_h$, respectively, at electron energy 0.20 hartree. These squared energy-normalized dipole amplitudes precede photon, spin and orientation prefactors. The factor of 5.14 shows the sensitivity of this weak transition to truncated wavefunction corrections and cancellation between signed contributions. The correction norms diagnose the expansion; they provide no rigorous bound on its neglected remainder. The SI specifies this quartic Hamiltonian and its finite-domain controls; Fig.~\ref{fig:internal} compares the harmonic and analytic internal-coordinate models.

\subsection{Fixed-photon water cross sections and angular signatures}
\begin{figure}[t]
\includegraphics[width=\linewidth]{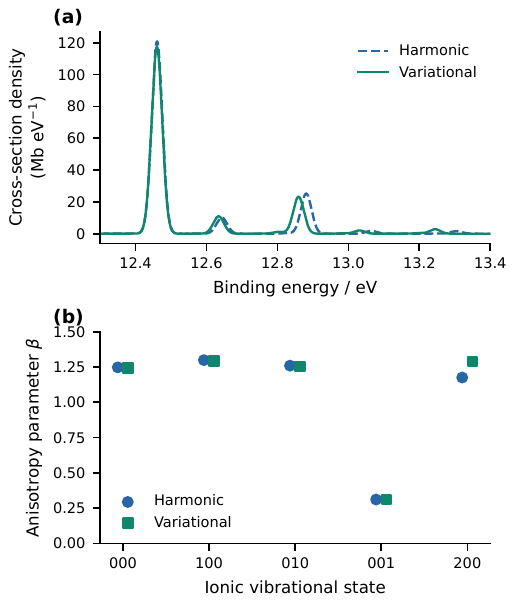}
\caption{Water at fixed photon energy: (a) first-channel cross-section density with illustrative Gaussian width $\sigma=0.015$~eV; (b) line-resolved anisotropy. Both ionic spin components are included.}
\label{fig:fixedphoton}
\end{figure}

Figure~\ref{fig:fixedphoton} presents the first-ionization-channel spectrum at photon energy 21.218~eV. Each ionic level is evaluated at its own outgoing-electron energy. The calculated origins are 12.46060~eV for the harmonic model and 12.46015~eV for ZAFF/DVR, with no adjustment to a measured ionization threshold. The variational origin is 0.161~eV below the evaluated experimental ionization energy of $12.621\pm0.002$~eV.\cite{NISTWater} This absolute-origin comparison tests the adopted electronic energy gap, separately from the ionic spacings in Table~\ref{tab:water}. The adopted continuum table spans 0.28--0.34 hartree in steps of 0.005 hartree.
Table~\ref{tab:fixedphoton} retains state assignments and separates the spectral change from the angular response. The first-channel cross section summed over the twelve retained variational ionic states is 5.9433~Mb; the twenty-one harmonic states in the same electron-energy domain give 6.0261~Mb. The two state sets are identified explicitly: the variational states carry approximately 99.1\% of their fixed-electron-energy sum rule, and the harmonic subset carries more than 99.94\%. These fractions refer to the corresponding fixed-energy nuclear checks, not an exact sum rule for an energy-dependent operator.
\begin{table}[t]
\caption{Assigned first-channel partial cross sections at 21.218~eV (Mb). The final column gives the ZAFF/DVR angular anisotropy.}
\label{tab:fixedphoton}
\begin{ruledtabular}
\begin{tabular}{lrrr}
Band & Harmonic & ZAFF/DVR & $\beta$\\\hline
000 & 4.54412 & 4.41479 & 1.241\\
100 & 0.38368 & 0.41104 & 1.293\\
010 & 0.94483 & 0.87097 & 1.253\\
001 & 0.00482 & 0.00437 & 0.311\\
200 & 0.01113 & 0.04398 & 1.289\\
\end{tabular}
\end{ruledtabular}
\end{table}
The antisymmetric fundamental is activated entirely by the linear HT term and has $\beta=0.311$, substantially below the origin value of 1.241. Its angular response adds an observable distinction to its weak intensity, reflecting interference among outgoing partial waves.\cite{Cooper1968,Reid2003} The variational bending-overtone cross section is 3.95 times the harmonic value, while the sum over retained lines changes much less.

Holding the continuum at the origin-line electron energy instead gives 5.9023~Mb for the retained variational spectrum. Transition-specific energy conservation therefore changes this sum by 0.694\%. Independent intermediate-energy calculations test the interpolation. Refining the electron-energy mesh changes retained line cross sections by at most 0.00822\%; the SI gives the mesh and nuclear-moment controls.

For the vibrationally integrated first ionic channel at 21.2~eV, Truesdale \emph{et al.}\cite{Truesdale1982} report $\sigma=5.77\pm0.14$~Mb and $\beta=0.99\pm0.08$. At this same photon energy the retained variational lines give 5.946~Mb and $\bar\beta=\sum_f\sigma_f\beta_f/\sum_f\sigma_f=1.246$: differences of 3.1\% and 0.26, respectively. The harmonic values are 6.029~Mb and 1.252. The static local model therefore describes the integrated intensity more closely than the angular distribution. This physical comparison is separate from numerical equivalence to a reference continuum solver. The line-resolved prediction $\beta(001)=0.311$ is distinct from the measured band average.

\subsection{Water angular-space and derivative controls}
Table~\ref{tab:water-physical-convergence} tests physical observables at the same photon energy as Fig.~\ref{fig:fixedphoton}, keeping the variational nuclear moments and radial representation fixed. The $l=6\to7$ change in the 001/origin ratio is 1.78\%, compared with 0.0063\% in the summed cross section. The anisotropy changes by $-0.00505$. At $l=7$, halving the central-difference step changes this ratio by 0.00039\% and $\beta(001)$ by $4.2\times10^{-7}$. The five-point result agrees with the finer central stencil. This separates derivative-step stability from angular-space sensitivity of the weak band.
\begin{table}[tbp]
\caption{Water at 21.218~eV with the same 12-state variational model. $\Sigma$ is the retained first-channel cross section; $r=100I_{001}/I_{000}$. Steps are dimensionless.}
\label{tab:water-physical-convergence}
\begin{ruledtabular}
\begin{tabular}{clrrr}
$l_{\max}$ & Stencil & $\Sigma$/Mb & $r$ (\%) & $\beta(001)$\\\hline
6 & $\pm0.02$ & 5.9429083 & 0.0971659 & 0.3156949\\
7 & $\pm0.02$ & 5.9432843 & 0.0988943 & 0.3106469\\
7 & $\pm0.01$ & 5.9432843 & 0.0988947 & 0.3106465\\
7 & Five-point & 5.9432844 & 0.0988948 & 0.3106463\\
\end{tabular}
\end{ruledtabular}
\end{table}

\subsection{Energy dependence and the shape of the removal orbital}
\begin{figure}[tbp]
\includegraphics[width=\linewidth]{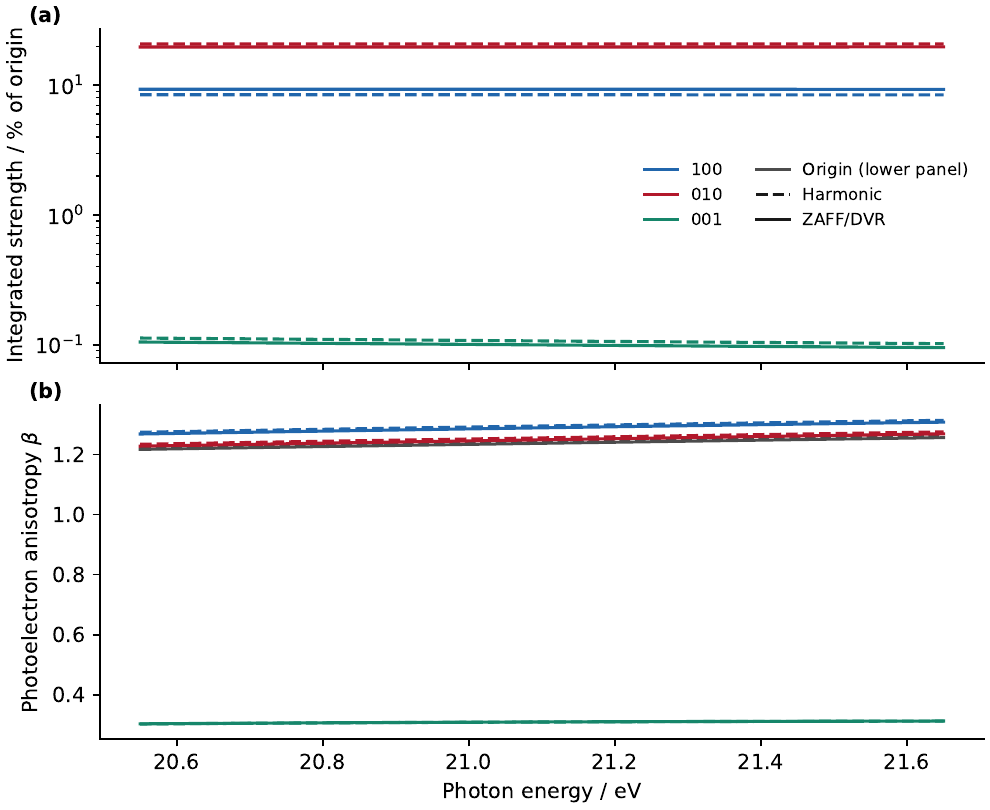}
\caption{Water fundamentals versus photon energy: (a) integrated intensity relative to the origin; (b) photoelectron anisotropy. Dashed lines: harmonic; solid lines: ZAFF/DVR.}
\label{fig:water-energy-bands}
\end{figure}

Figure~\ref{fig:water-energy-bands} follows three water fundamentals across
20.55--21.65~eV, using the existing complex-amplitude table and assigning a
different electron energy to each vibrational line. The antisymmetric
fundamental remains a sensitive probe of the continuum: its variational
relative intensity decreases from 0.1052\% to 0.0952\% of the origin across
this interval, while its anisotropy lies between 0.303 and 0.313. The origin
has $\beta=1.217$--1.256. Harmonic and variational nuclear states therefore
sample a common electronic operator with distinct intensity and angular
responses.

\begin{figure}[tbp]
\includegraphics[width=\linewidth]{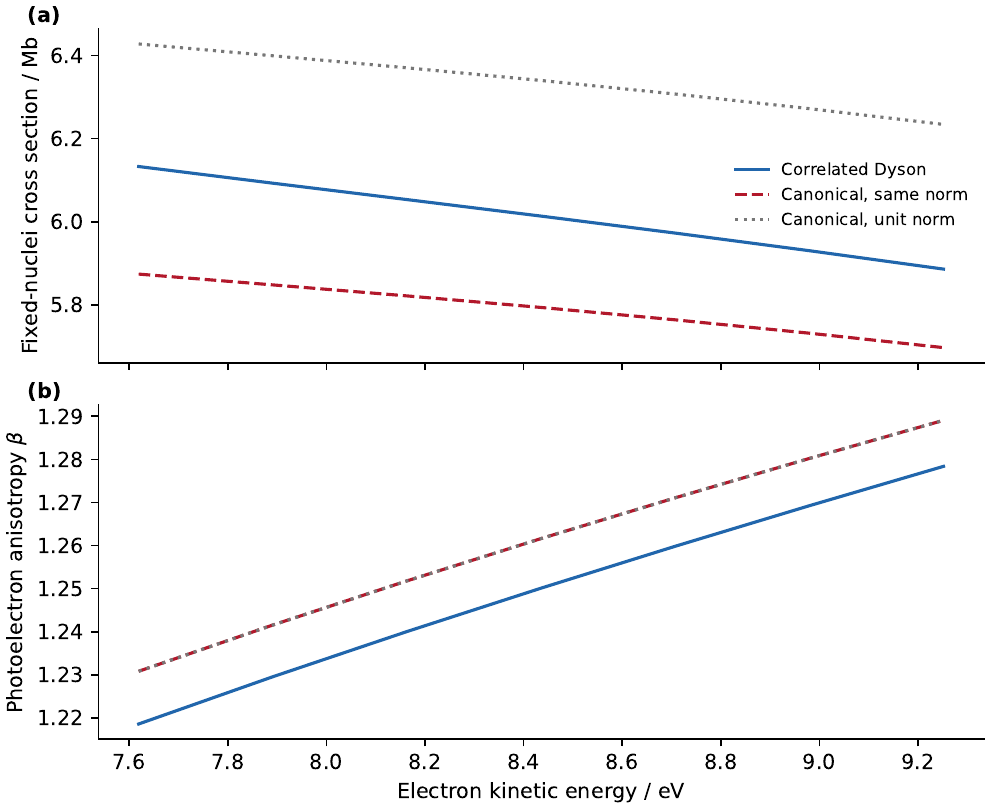}
\caption{Water fixed-nuclei source comparison in the same continuum: (a) cross section; (b) anisotropy. The norm-matched canonical source isolates the effect of orbital shape.}
\label{fig:water-orbital-shape}
\end{figure}

A fixed-nuclei comparison separates the shape of the correlated Dyson orbital
from its norm. The closest occupied canonical orbital has normalized overlap
0.999458 with the NR2 Dyson and is scaled to the same squared norm,
$Z=0.913856$. Both sources are projected into the identical $l\leq7$
continuum, with the same density, potential, boundary conditions and removal
energy. Across electron energies 0.28--0.34 hartree, the correlated source
gives cross sections 3.32--4.41\% larger than the norm-matched canonical
source, while $\beta$ decreases by 0.0107--0.0122. Figure~\ref{fig:water-orbital-shape}
shows that a large orbital overlap still permits observable changes in
photoionization. A unit-norm canonical source provides a separate control:
rescaling it by $\sqrt{Z}$ multiplies its cross section by $Z$ and leaves
$\beta$ unchanged. These results connect the complete signed-vector treatment
to experimental evidence that orbital mixing affects dynamical observables
more strongly than pole strengths.\cite{Schio2026}

\begin{figure*}[tp]
\includegraphics[width=\textwidth]{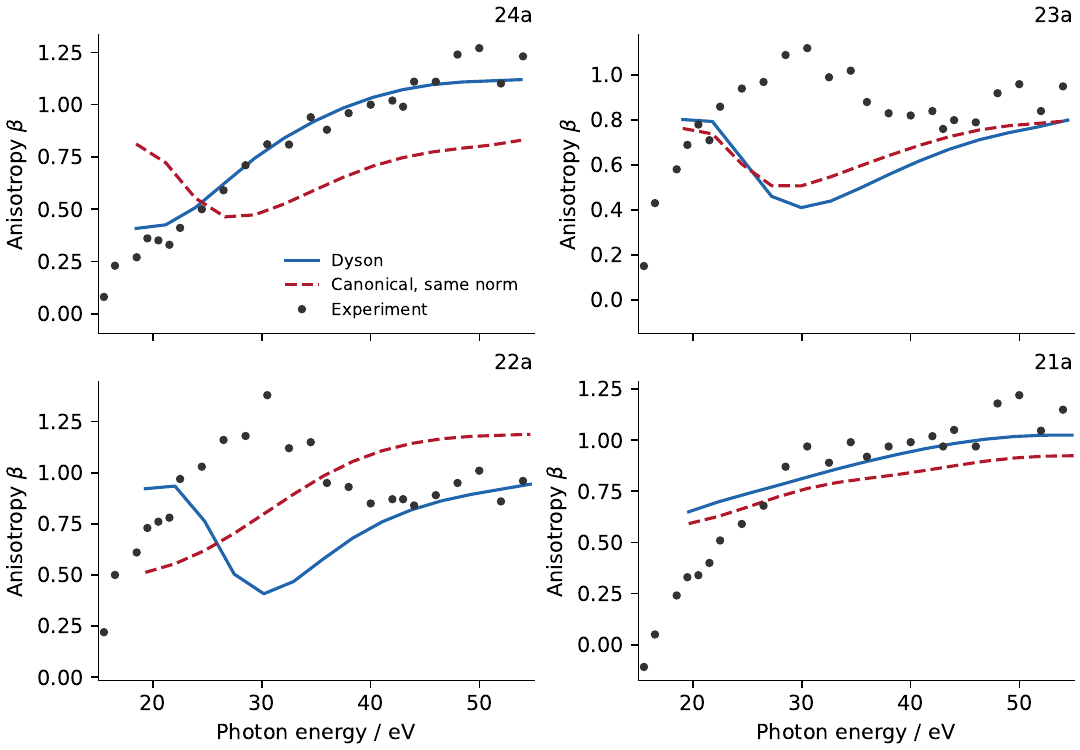}
\caption{Epichlorohydrin outer-valence anisotropies. Solid: full NR2 Dyson;
dashed: norm-matched canonical source in the same static field for g-II.
Dots: experimental conformer mixture, recovered from the published vector
figure.\cite{Schio2026} The SI specifies the continuum space and integration
sensitivity and compares relative partial strengths.}
\label{fig:epichlorohydrin-beta}
\end{figure*}

Epichlorohydrin supplies a stronger orbital-mixing example. A single
NR2/aug-cc-pVDZ calculation at the published g-II geometry, the conformer
with a reported population of 68\%,\cite{Schio2026} gives the four
outer-valence channels in Table~\ref{tab:epichlorohydrin-sources}.
The channel labels follow the measured energy ordering.
The squared Dyson norms remain close to 0.9, while two normalized
Dyson--canonical overlaps fall below 0.72. At an electron kinetic energy
of 32.65~eV, replacing the norm-matched canonical source by its Dyson
changes the first and third channel cross sections by factors of about
2.5 and 0.33. The Cl-centred radial support extends to 2 bohr;
the other atomic supports have radius 0.8 bohr. These source-shape contrasts
persist across the declared angular refinements. The SI reports the angular sensitivity of each
cross section and anisotropy. This fixed-conformer comparison uses a
common static local field throughout. Figure~\ref{fig:epichlorohydrin-beta}
compares the four anisotropy profiles with the measured room-temperature
mixture. The experimental 23a and 22a bands show a maximum near
30--32~eV and a minimum near 43~eV. The static g-II profiles place their
minima at lower photon energy; the full Dyson substantially changes the
24a and 22a profiles. The SI gives relative partial strengths, the numerical
integration controls and the distinction from the published multiconformer
equation-of-motion ionization-potential coupled-cluster singles and
doubles (EOM-IP-CCSD) calculations with electronic screening.\cite{Schio2026}
\begin{table}[tbp]
\caption{Epichlorohydrin g-II removal channels. $Z$: squared Dyson norm;
$s$: normalized overlap with the closest occupied canonical orbital;
$R$: Dyson/canonical cross-section ratio at equal norm and electron
energy 32.65~eV. The interval spans $l_{\max}=13$ and 15 with a 2-bohr Cl support.}
\label{tab:epichlorohydrin-sources}
\begin{ruledtabular}
\begin{tabular}{crrrr}
Channel & Removal energy/eV & $Z$ & $s$ & $R$\\\hline
24a & 10.298 & 0.8834 & 0.7153 & 2.490--2.520\\
23a & 10.907 & 0.9037 & 0.9829 & 0.852--0.854\\
22a & 11.144 & 0.8974 & 0.6869 & 0.321--0.337\\
21a & 11.451 & 0.8932 & 0.9526 & 1.048\\
\end{tabular}
\end{ruledtabular}
\end{table}

\subsection{Furan: coupled nuclear motion and weak photoelectron bands}
Furan extends the calculation to a nine-atom molecule with 21 vibrational
coordinates. Its first photoelectron band contains a dominant origin,
several ring-mode progressions and weak features that probe departures from
the Condon approximation. The experimental band structure and its
nonadiabatic interpretation provide complementary references for this
problem.\cite{Derrick1971,Trofimov1998} Jet-cooled resonance-enhanced multiphoton ionization (REMPI) measurements assign a neutral $3d$ Rydberg progression and connect its vibrational structure to the first photoelectron band.\cite{Philis2007} These assignments provide complementary context for identifying ring-mode progressions.

The harmonic fields of the neutral and cation use
revDSD-PBEP86-D3/jun-cc-pVTZ. A composite nuclear Hamiltonian replaces the
full quadratic matrix and residual forces of the transported
Becke three-parameter hybrid with Lee--Yang--Parr correlation (B3LYP)/6-31+G(d) quartic field\cite{Becke1993,LYP1988} by these higher-level values. Both surfaces
retain explicit linear terms and all acquired higher interactions.
Selected vibrational configuration interaction (VCI) concentrates the nuclear basis on coupled product states. Modern selected-VCI approaches demonstrate the value of this strategy for many-mode anharmonic systems.\cite{Fetherolf2021,Tran2023} Here the two surfaces use 982 neutral and
818 ionic product functions, with all
21 modes and their complete Duschinsky transformation. Separate enlargement
of the two spaces stabilizes the broadened low-energy window: resolved
maxima move by at most 2~cm$^{-1}$, and integrated band groups change by
at most 0.338\% in the final ionic-space control. The SI reports the
independent neutral-space and TI/TD checks. The window extends from $-150$ to
1800~cm$^{-1}$ relative to the harmonic origin, with Gaussian broadening
$\sigma=50$~cm$^{-1}$. Convergence is assessed within this window.

\begin{figure*}[t]
\centering
\includegraphics[width=.88\textwidth]{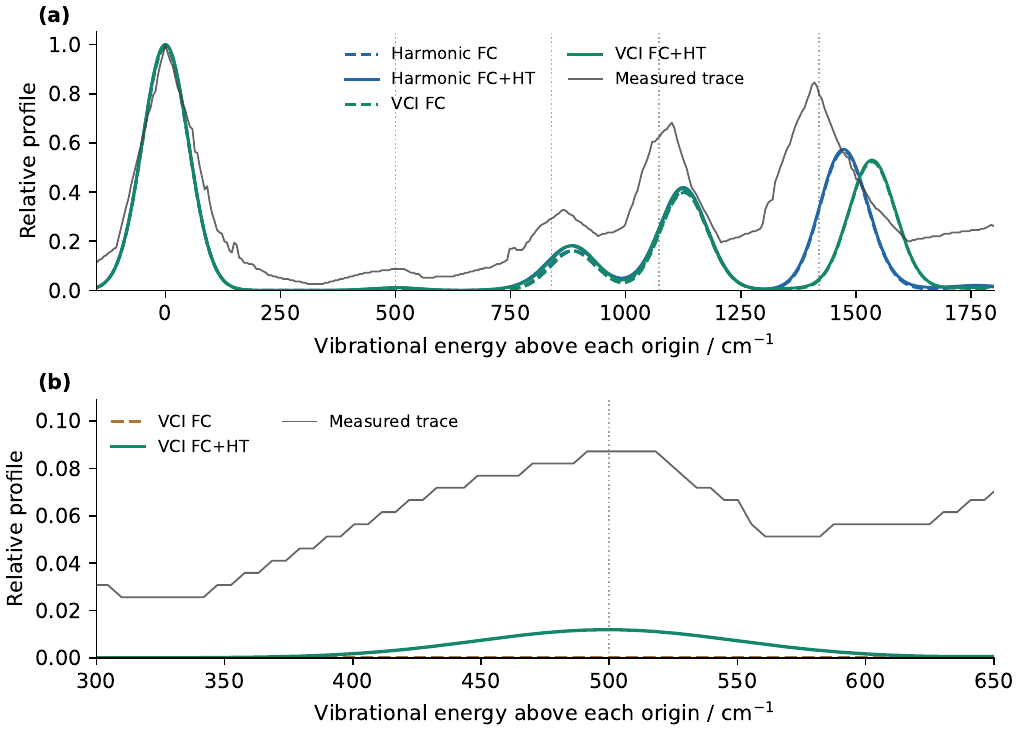}
\caption{Furan at 21.218~eV: (a) harmonic and VCI profiles with the measured trace; (b) continuum-induced intensity near 500~cm$^{-1}$. Profiles use their own origins and origin-peak normalization; $\sigma=50$~cm$^{-1}$.}
\label{fig:furan-spectrum}
\end{figure*}

Figure~\ref{fig:furan-spectrum} compares the harmonic and VCI profiles at 21.218~eV with the measured first photoelectron band.\cite{Derrick1971,Trofimov1998} The profiles use the final $l\leq11$ central continuum with compact atomic functions and the same electronic operator for both nuclear models. Each calculated profile is referred to its own origin to compare vibrational structure; the absolute origin is assessed below. The ring-mode progressions dominate the envelope, while panel~(b) compares the weak out-of-plane region directly with the measured trace, including its background.

Here $j$ orders the ionic harmonic modes by frequency, and $w_j$ is the
weight of the corresponding one-quantum product function in a VCI state. The relative integrated cross section $I_j/I_0$ compares each fundamental with the origin at the same photon energy.
Table~\ref{tab:furan-nuclear} illustrates the state-resolved nuclear
information retained by the calculation. The two lowest out-of-plane
fundamentals lie at 491 and 509~cm$^{-1}$; reflection makes their FC overlaps
vanish, so their photoelectron intensities probe the coordinate-dependent
continuum operator. The measured weak feature near
500~cm$^{-1}$\cite{Trofimov1998} makes this region particularly informative.
The variational state near 1086~cm$^{-1}$ has a leading product weight of
only 69\%, illustrating the mixing retained in the coupled treatment. The two lowest fundamentals have relative integrated cross sections of 0.634\% and 0.578\% and form the weak envelope around 500~cm$^{-1}$. Their FC contributions vanish, whereas the dominant ring-mode bands retain substantial FC strength. This distinguishes continuum-induced activation from redistribution by nuclear anharmonicity. The measured progression also tests the potential. For mode 16, the positive anharmonic shift of 60.6~cm$^{-1}$ moves the harmonic band at 1471.1 to 1531.7~cm$^{-1}$, increasing its difference from the observed feature near 1420~cm$^{-1}$ from about 51 to 112~cm$^{-1}$. Thus the composite anharmonic field improves the description of coupled nuclear states without improving this particular experimental spacing. Its transported B3LYP quartic terms and revDSD quadratic terms define a specific approximate surface; a controlled change of that surface would be needed to isolate the origin of the shift.

The calculated relative envelope underestimates measured features near 850, 1080 and 1420~cm$^{-1}$. Near 500~cm$^{-1}$, the calculated HT profile reaches about 0.012 relative to the origin, below the measured trace including background. The calculation identifies an allowed contribution to this region, not its full experimental intensity. Trofimov, K\"oppel and Schirmer attribute intensity borrowing near 799 and 1565~cm$^{-1}$ to coupling of the $^2A_2$ and $^2B_1$ ionic states through in-plane modes; their table lists the weak 500~cm$^{-1}$ feature without a corresponding theoretical assignment.\cite{Trofimov1998} Our out-of-plane HT activation and that two-state mechanism therefore concern distinct assignments. Combining both mechanisms requires consistent electronic states: adiabatic amplitude derivatives can already contain state mixing, which an explicit diabatic coupling must account for.

\begin{table}[b]
\caption{Furan cation fundamentals (cm$^{-1}$), product weights and relative integrated cross sections at 21.218~eV.}
\label{tab:furan-nuclear}
\begin{ruledtabular}
\begin{tabular}{rrrrr}
$j$ & Harmonic & VCI & $w_j$ & $I_j/I_0$ (\%)\\
\hline
1 & 485.1 & 490.6 & 0.957 & 0.634\\
2 & 509.4 & 508.7 & 0.956 & 0.578\\
6 & 882.2 & 886.8 & 0.955 & 16.04\\
10 & 1073.1 & 1086.0 & 0.691 & 1.76\\
12 & 1125.8 & 1126.9 & 0.923 & 37.91\\
16 & 1471.1 & 1531.7 & 0.811 & 40.60\\
\end{tabular}
\end{ruledtabular}
\end{table}

The electronic origin uses the complete signed NR2/jun-cc-pVTZ Dyson
orbital and its calculated vertical pole. Combining it with the nuclear
energies gives a VCI origin of 8.783~eV, compared with the measured
adiabatic ionization energy of $8.8863\pm0.0002$~eV.\cite{Yang2006}
Vibrational spacings and the absolute electronic origin therefore provide
separate comparisons. The complete continuum comparison uses $l_{\max}=9$ and 11 with atomic
Lebedev orders 17 and 29, respectively. It changes absolute FC+HT profiles
by 2.47--2.59\% at electron energies 0.40--0.50 hartree, with unchanged
resolved maxima. The final reference uses $l_{\max}=11$; a separate
order-35 central quadrature control changes the equilibrium cross sections
by less than 0.000083\%. The SI reports these controls independently of the
nuclear-space enlargement.

\subsection{Observable convergence and continuum sensitivity}
Table~\ref{tab:furan-ht-angular} separates the continuum representation,
central quadrature and nuclear-space controls. In the first two rows,
the representation labels give $l_{\max}$/atomic Lebedev order, and the
strength changes refer to $r_j=I_j/I_0$. The centre-only control refines
quadrature at fixed $l_{\max}=11$ and reports the equilibrium cross section
and anisotropy. The nuclear rows give the largest absolute-profile
$L^1$ change over the three electron energies, normalized to the reference
window area. Resolved nuclear-space peak changes remain at most
2~cm$^{-1}$. The SI gives absolute band strengths and anisotropies.
\begin{table}[tbp]
\caption{Furan sensitivity to continuum representation, central quadrature
and nuclear space. The final reference uses $l_{\max}=11$ and Lebedev order 29.}
\label{tab:furan-ht-angular}
\begin{ruledtabular}
\small
\setlength{\tabcolsep}{2.5pt}
\begin{tabular}{llrr}
Refinement & Observable & Change (\%) & $|\Delta\beta|$\\\hline
$9/17\to11/29$ & $r_1$ & $-4.22$ & 0.03352\\
 & $r_2$ & $-5.02$ & 0.07127\\
Centre $29\to35$ & $\sigma_{\rm eq}$ & $<0.000083$ & $<1.32\!\times\!10^{-6}$\\
Neutral $950\to982$ & Profile $L^1$ & 0.136 & ---\\
Ionic $786\to818$ & Profile $L^1$ & 0.788 & ---\\
\end{tabular}
\end{ruledtabular}
\end{table}

\subsection{Efficient numerical representations}
Matched-accuracy H$_2^+$ FE-DVR and B-spline calculations take 0.818 and 1.921~s for three energies, respectively. Compact-support assembly retains all couplings; the SI specifies basis dimensions and separates assembly costs.

Continuum dipoles are evaluated by an exact transposed-source solve that avoids constructing unneeded scattering matrices. Direct shell-pair Coulomb contractions, shared angular projections and fill-reducing ordering retain all operators and channels. The fixed sparse structure and column ordering are reused across energies, with a fresh numerical factorization at each energy. On a 5632-function water problem, matched one-thread continuum solves take 12.35 versus 0.643~s, including factorization and excluding field preparation. The SI distinguishes these costs and documents molecular amplitude checks.

Repeated spectra reuse distinct physical ingredients at their natural level. When ionic channels use the same local field, its preparation and the factorization at each electron energy are shared. Nuclear transition overlaps are retained by state assignment and reused across photon energies and electronic operators; each operator retains its own TI selection and intensity-recovery test. These separations make energy-dependent band assignments and channel comparisons practical. With the field and dipole integrals prepared, the four-channel epichlorohydrin Dyson/canonical outgoing solution costs 7.1\% more than the one-channel solution. The SI gives the complete timing conditions, nuclear-overlap checks and bounded-memory line broadening.

A matched 15-energy CO response calculation gives a factor of 18.6 for spectral evaluation over projected sparse solves, including preparation and retaining the same complete operators and boundary conditions. The SI reports dimensions, hardware and repeated timings. Charge-preserving density fitting\cite{Whitten1973,Dunlap1979} accelerates Hartree-field evaluation for larger AO spaces, with exact Gaussian fields as reference.

\subsection{Relation to established continuum methods}
Table~\ref{tab:continuum-methods} compares established molecular continuum tools by their treatment of the departing electron and electronic target. Numerical timing comparisons require the same Hamiltonian, continuum space, boundary conditions and observable.

\begin{table}[tbp]
\caption{Representative electronic and continuum descriptions.}
\label{tab:continuum-methods}
\begin{ruledtabular}
\begin{tabular}{p{2.0cm}p{5.6cm}}
Tool & Description\\\hline
TIRESIA\cite{Tiresia2024,Tenorio2022} & Dyson orbitals and local density-functional scattering; multicentre B-splines.\\[3pt]
ezDyson\cite{ezSpectra2022} & Correlated Dyson orbitals with plane or Coulomb waves and orientation averaging.\\[3pt]
ePolyScat\cite{ePolyScat1999,ePolyScatDocs} & Static-exchange scattering and approximate polarization; single-centre harmonics and radial grids.\\[3pt]
UKRmol+\cite{UKRmol2020} & Multielectron R matrix with coupled electronic channels; Gaussian or mixed Gaussian/B-spline functions.\\[3pt]
SPARKS & Signed Dyson orbitals, local scattering and projected response; B-spline or FE-DVR spaces coupled to harmonic and anharmonic nuclear states.\\
\end{tabular}
\end{ruledtabular}
\end{table}

The independent comparisons test complex amplitudes, cross sections and angular anisotropies. A matched CO response calculation also separates electronic screening from continuum propagation. The SI gives the corresponding potential, susceptibility and screened-amplitude comparisons. The practical contributions are interchangeable radial representations, efficient energy scans and a common complex electronic operator throughout the nuclear hierarchy.

The same complex Cartesian amplitudes support molecular-frame angular distributions for linear, circular, elliptical and mixed polarization. A positive, unit-trace polarization density matrix specifies the light; atomic patterns and angular sum rules test the contraction. The shared convention connects nuclear assignments, absolute strengths and directional observables. A self-contained distribution reproduces the calculations from electronic-structure data.

Recent molecular applications emphasize complementary observables. For OCS,
continuum calculations combined with molecular alignment resolve angular
patterns and their sensitivity to nuclear geometry.\cite{Woodhouse2025}
For glycine, TIRESIA ionization probabilities initialize quantum and mixed
quantum--classical simulations of proton transfer.\cite{Kety2025}
The water delay calculation includes correlated electronic channels and
one- and two-photon couplings, with harmonic vertical-Hessian nuclear states
and FC transition matrix elements.\cite{Pranjal2025} The present water and
furan applications instead resolve how anharmonic nuclear states and the
coordinate dependence of the continuum redistribute single-photon band
strengths and angular anisotropies. The full-dimensional anharmonic water
study of Engin and co-workers supplies the earlier direct nuclear-continuum
reference.\cite{Engin2019}

\subsection{Complementary spectroscopic signatures}
\begin{figure*}[t]
\includegraphics[width=\textwidth]{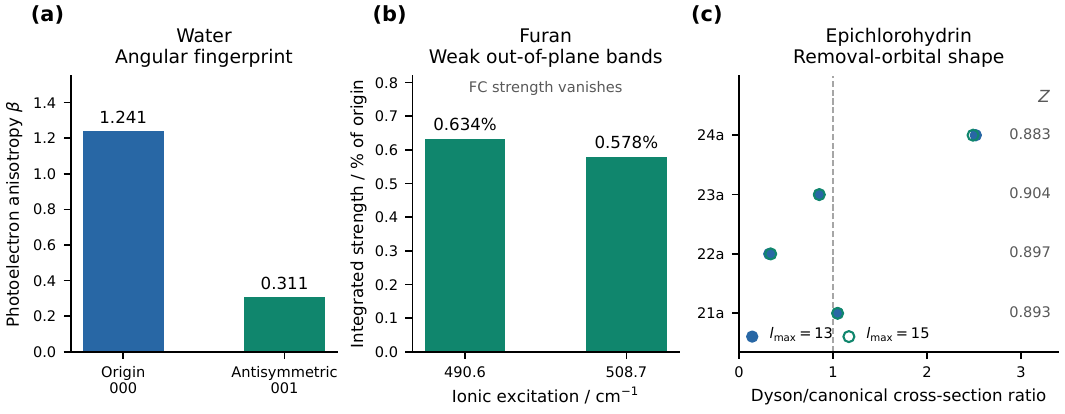}
\caption{Complementary molecular signatures: (a) water origin and antisymmetric-band anisotropies; (b) relative integrated strengths of the two lowest furan out-of-plane bands, both FC-dark; (c) epichlorohydrin Dyson/canonical cross-section ratios at equal norm, with squared Dyson norms $Z$. Water and furan use 21.218~eV photons; panel (c) uses electron energy 1.2 hartree. The two angular spaces in (c) are numerical comparisons.}
\label{fig:spectroscopic-summary}
\end{figure*}

Figure~\ref{fig:spectroscopic-summary} brings together three observables that separate the electronic and nuclear contributions. Water reveals continuum-induced activation through a distinct angular signature; the furan bands identify weak coupled nuclear states; epichlorohydrin isolates removal-orbital shape at a common norm. Together they connect the same signed-amplitude description to band assignments and channel-resolved intensities.

\section{Conclusions}
Photoelectron band positions and intensities respond to distinct parts of the ionization process. Nuclear reorganization sets the vibrational progression, while the coordinate dependence and phase of the electronic continuum amplitude activate weak bands and determine their angular signatures. SPARKS combines these contributions with transition-specific energy conservation and state-resolved assignments.

Water makes their separate roles quantitative. The internal-coordinate variational bending spacing is 1403.4~cm$^{-1}$, compared with 1471.2~cm$^{-1}$ in the harmonic model and 1406.6~cm$^{-1}$ experimentally. At 21.218~eV, the FC-dark antisymmetric fundamental has a partial cross section of 0.00437~Mb and an anisotropy of 0.311, compared with 1.241 for the origin. Thus a weak band can carry an angular signature distinct from the dominant progression. The band-integrated experimental comparison at 21.2~eV gives a 3.1\% cross-section difference and an anisotropy difference of 0.26, distinguishing physical model accuracy from numerical convergence.

Furan extends the nuclear contraction to a coupled 21-mode VCI window. Its two lowest out-of-plane fundamentals occur at 490.6 and 508.7~cm$^{-1}$ and acquire their intensity through the continuum derivative. In the reported atomic-enriched $l\leq11$ calculation, their integrated strengths are 0.634\% and 0.578\% of the origin. The comparison of harmonic and variational profiles separates these electronic activation effects from shifts and mixing induced by the nuclear Hamiltonian.

The epichlorohydrin comparison extends this separation to four ionic states: similar removal probabilities coexist with strongly different orbital shapes. Channel-resolved strengths and anisotropies expose those differences directly, complementing the vibrational sensitivity demonstrated by water and furan.

Independent continuum formulations and established molecular references test normalization, phases and electronic screening. The common transition operator then connects these electronic controls to harmonic, perturbative and variational nuclear descriptions. The resulting observables link band assignments, anharmonic spacings and angular distributions within the same physical model.

\section*{Conflicts of interest}
There are no conflicts of interest to declare.
\section*{Data and code availability}
SPARKS is available as a self-contained source distribution at \href{https://github.com/yogibubu/SPARKS-standalone}{github.com/yogibubu/SPARKS-standalone}. Release 0.1.0rc8 supplies the source code, tests, installer, examples and dependency inventories; platform-specific offline dependencies are provided as a separate release asset.\cite{SPARKSRelease} The public distribution is independent of the development repository and does not require TIRESIA.

Portable continuum amplitudes, nuclear transition moments, figure data, convergence records and reproduction scripts are supplied in the accompanying numerical-data archive. Its reproduction guide identifies the definitive data used in each molecular example. Nuclear basis refinement evaluates the specified analytic potentials with a fixed electronic acquisition set. Licensed electronic-structure executables are obtained separately from their providers.

\end{document}